\documentclass[prl]{revtex4}

\usepackage{graphicx}
\usepackage{dcolumn}
\usepackage{bm}

\begin{document}
\title{Grover Energy Transfer at Relativistic Speeds} 
\author{Juan Carlos Garc\'ia-Escart\'in${}^{1,2}$}
\email{juagar@tel.uva.es}
\author{Pedro Chamorro-Posada${}^1$}%
\affiliation{%
${}^{1}$Departamento Teor\'ia de la Se\~{n}al y Comunicaciones e Ingenier\'ia Telem\'atica. Universidad de Valladolid.\\
ETSI Telecomunicaci\'on, Campus Miguel Delibes. Camino del Cementerio s/n, 47011 Valladolid, Spain.
}%
\affiliation{
${}^{2}$Centre for Quantum Computation, Department of Applied Mathematics and Theoretical Physics, University of Cambridge, Wilberforce Road, Cambridge CB3 0WA, United Kingdom.
}

\date{\today}
\begin{abstract}
Grover's algorithm for quantum search can also be applied to classical energy transfer. The procedure takes a system in which the total energy is equally distributed among $N$ subsystems and transfers most of the it to one marked subsystem. We show that in a relativistic setting the efficiency of this procedure can be improved. We will consider the transfer of relativistic kinetic energy in a series of elastic collisions. In this case, the number of steps of the energy transfer procedure approaches 1 as the initial velocities of the objects become closer to the speed of light.  This is a consequence of introducing non-linearities in the procedure. However, the maximum attainable transfer will depend on the particular combination of speed and number of objects. In the procedure, we will use $N$ elements, like in the classical case, instead of the $log_2(N)$ states of the quantum algorithm.
\end{abstract}

\pacs{Valid PACS appear here}
\maketitle

\section{Introduction}
Physics gives new ways to analyse information transmission and computation. The traditional theories of information and communication implicitly assume an underlying classical world, with no regard to modern physical theories. However, the recent physical analysis of information has considerably enriched both fields. Taking quantum mechanics into account in communications has led to quantum cryptography \cite{BB84,Eke91} and the applications to information processing include efficient algorithms for factoring \cite{Sho97} among others \cite{NC00}. 

But, while quantum information has become an active field of research, there are few results that incorporate the principles of the theory of relativity into communications and information processing. There are already promising studies that include relativistic constraints into new communication protocols \cite{BHK05,BHP09}. In particular, relativistic bit commitment protocols offer results with no classical or quantum counterpart \cite{Ken99,HK04}. In spite of these advances, there has been no significant work on the possibility of relativistic computing. 

In this paper, we take a variation of Grover's algorithm for quantum search and translate it into a relativistic setting. This variation can be though of as an effective procedure for energy transfer between $N$ classical objects. In the classical case, the algorithm needs a number of steps proportional to $\sqrt{N}$ to transfer the energy to a marked object. We show that special relativity predicts a smaller number of steps, which can be reduced even to a single step when the speed of light, $c$, is approached.

\section{Grover's algorithm}
The basic form of Grover's algorithm for quantum search gives a procedure to stand out a marked element out of $N$ possibilities \cite{Gro97}. The algorithm relies on an oracle that can detect and invert the phase of a particular target state $|t\rangle$. The starting point is a uniform superposition $|s\rangle$ of all the possible states. With a sequence of oracle calls and diffusion operators that redistribute the probability amplitudes of a state among the others, we can gradually transfer the probability amplitude from the non-marked states to the target. After a number of iterations of the order of $\sqrt{N}$, there is a high probability of finding the desired state in a measurement.

The evolution can be more easily understood if we define a state $|t_{\perp}\rangle$ that collects all the non-marked states from the original superposition. This state is orthogonal to $|t\rangle$ and close to $|s\rangle$ in the original Hilbert space. 
If we observe the evolution of $|s\rangle$ in the two-dimensional space spanned by $|t\rangle$ and $|t_{\perp}\rangle$, Grover's algorithm reduces to a series of rotations that take the state from $|s\rangle$ to $|t\rangle$ \cite{FG98}.

A similar procedure can be defined to transfer energy between classical oscillators \cite{GS02}. In the initial state, the total energy of the system is equally distributed among $N$ oscillators. During the Grover procedure, the energy of a marked oscillator will go from being of the order of $\frac{1}{N}$th of the total energy to being of the order of the total energy. Now, the diffusion operator redistributes energy instead of probability amplitude. The crucial difference is that in the classical case we can no longer use a quantum superposition. This results in an increase in the space complexity of the procedure. In the quantum case we could encode $N$ states in $log_2 N$ two-level systems. Now, we need $N$ oscillators. Apart from that, the time complexity of the procedure (the number of steps) is the same in both scenarios. 

Grover energy transfer can also be better studied by defining two different modes. The first one corresponds to the marked mode to which we want to transfer the energy of the global system of $N$ oscillators. The second mode is the collective mode of all the non-marked oscillators. 

We will discuss a simplified case based on elastic collisions \cite{ZL03}. The basic model will be the same for both classical and relativistic Grover energy transfer. We will consider $N$ bodies of normalized mass $1$ which, to simplify the study, will be grouped in two balls: a big ball of normalized mass $M=N-1$ and a small ball of normalized mass $1$. Both balls are moving to the right with the same initial speed $v_0$. Our algorithm will be designed to transfer all the kinetic energy of the composite system to the small ball. The transfer will occur in a series of elastic collisions. 

\begin{figure}
\includegraphics[width=\linewidth]{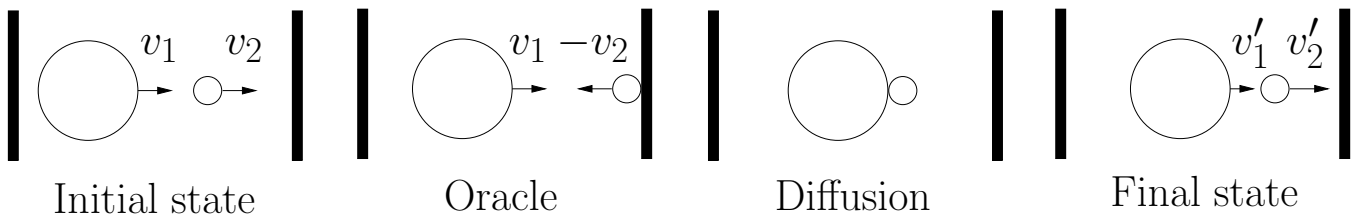}
\caption{\label{iteration} An iteration of the Grover energy transfer. In the initial state, the big ball moves to the right with velocity $v_1$ and the small one with velocity $v_2$. First, the small ball rebounds from the wall in an operation equivalent to the oracle sign shift of quantum search. After that, both balls collide. It is in the collision where the energy is transferred between the balls. This step is equivalent to the diffusion operation of Grover search. As a result of the collision, the balls move with velocities $v_1'$ and $v_2'$. This final state is the initial state of next iteration.}
\end{figure}

The situation is summed up in Figure \ref{iteration}. We begin with both balls moving to the right (with positive speeds $v_0$). To simplify things, we take the smaller ball to be a little ahead of the big one. If needed, this separation could be part of the oracle phase. The small ball will meet a rigid wall that will act as the oracle in Grover's algorithm. Now, instead of a $\pi$ phase shift, we have a change of sign in the velocity when the small ball rebounds. After the rebound, the balls will collide. Collisions play the role of the diffusion operator. They redistribute the energy into all the modes, in this case, the kinetic energy of the small and big balls. This will be the setting of both the classical and relativistic Grover energy transfer procedures. The small ball will start with one $N$th of the total kinetic energy and, at the end of the procedure, its fraction of the total kinetic energy will be close to one. We suppose that during the whole series of collisions, the big ball does not meet the wall and still moves to the right. This is not strictly necessary, but makes the analysis more direct. 

In the classical case, the evolution of the system after each collision is derived from conservation of energy and momentum. This evolution, when written in a matrix form, corresponds exactly to the evolution operator of quantum search and requires the same number of steps \cite{ZL03}. Notably, this does not depend on the initial velocity.

Here, we generalize the situation to relativistic speeds. Our analysis is based on special relativity and will pay no regard to any possible gravitational effects. All the given parameters are considered in the frame in which the wall is stationary.

We assume a series of elastic collisions. The evolution can be deduced from the conservation of four-momentum. We consider a simplified case in which there is only linear movement in the $x$ direction. All the velocities given will be taken to be in the $x$ direction and are normalized to the speed of light, using $v_i$ as a shorthand for $\frac{v_i^x}{c}$. In this case, conservation of four-momentum can be expressed with only two equations (for conservation of relativistic energy and $x$-momentum). Additionally, as we are dealing with elastic collisions, the rest mass of each ball will be conserved. The evolution after a collision is given by
\begin{eqnarray}
\label{consrele}
M\gamma_1+\gamma_2&=&M\gamma_1'+\gamma_2'\\
\label{consrelmom}
M\gamma_1v_1+\gamma_2v_2&=&M\gamma_1'v_1'+\gamma_2'v_2'
\end{eqnarray}
for the conservation of energy and $x$-momentum, respectively. In the equations, $\gamma_i$ represents the Lorentz factor $\frac{1}{\sqrt{1-v_i^2}}$, corresponding to velocity $v_i$. The left hand side of the equations define the four-momentum components before the collision and the right hand side the situation after the collision. Variables with subindices 1 refer to parameters of the big ball and those with subindices 2 correspond to the small ball. 

The resulting non-linear system of equations can be more easily solved by adding and subtracting both equations to give the equivalent system,
\begin{eqnarray}
M\alpha_1+\alpha_2&=&M\alpha_1'+\alpha_2'\\
\frac{M}{\alpha_1}+\frac{1}{\alpha_2}&=&\frac{M}{\alpha_1'}+\frac{1}{\alpha_2'},
\end{eqnarray}
where $\alpha_i=\sqrt{\frac{1+v_i}{1-v_i}}$. Now, we only need to solve the quadratic equation $A\alpha_2'^2+B\alpha_2'+C=0$, with $A=\frac{M}{\alpha_1}+\frac{1}{\alpha_2}$, $B=-\left[ M \right(\frac{\alpha_1}{\alpha_2}+\frac{\alpha_2}{\alpha_1}\left)+2\right]$ and $C=M\alpha_1+\alpha_2$. From the two possible solutions, we take the one that corresponds to the big ball still moving to the right and the small ball rebounding from it. With this value of $\alpha_2'$, we can recover $\alpha_1'=\alpha_1+\frac{1}{M}(\alpha_2-\alpha_2')$.  The final velocities can then be calculated from $v_i=\frac{\alpha_i^2-1}{\alpha_i^ 2+1}$. 

While we could study the energy transfer using velocity, we can understand the evolution better by looking into the relativistic kinetic energies of the balls. We have total energies $E_1=M\gamma_1$ and $E_2=\gamma_2$, but, as masses are conserved, the only energy that can be transferred is kinetic energy $K_1=M\gamma_1-M$ and $K_2=\gamma_2-1$. We are not interested in the particular values of these energies, but in their relationship to the total kinetic energy $K_T$. $K_T$ is conserved all throught the energy transfer procedure and can be deduced from the initial condition $K_T=K_1^{ini}+K_2^{ini}=N\gamma_0-N$.

\subsection{Results}
\subsection{Reduction in the number of steps}
\begin{figure}
\includegraphics[width=\linewidth]{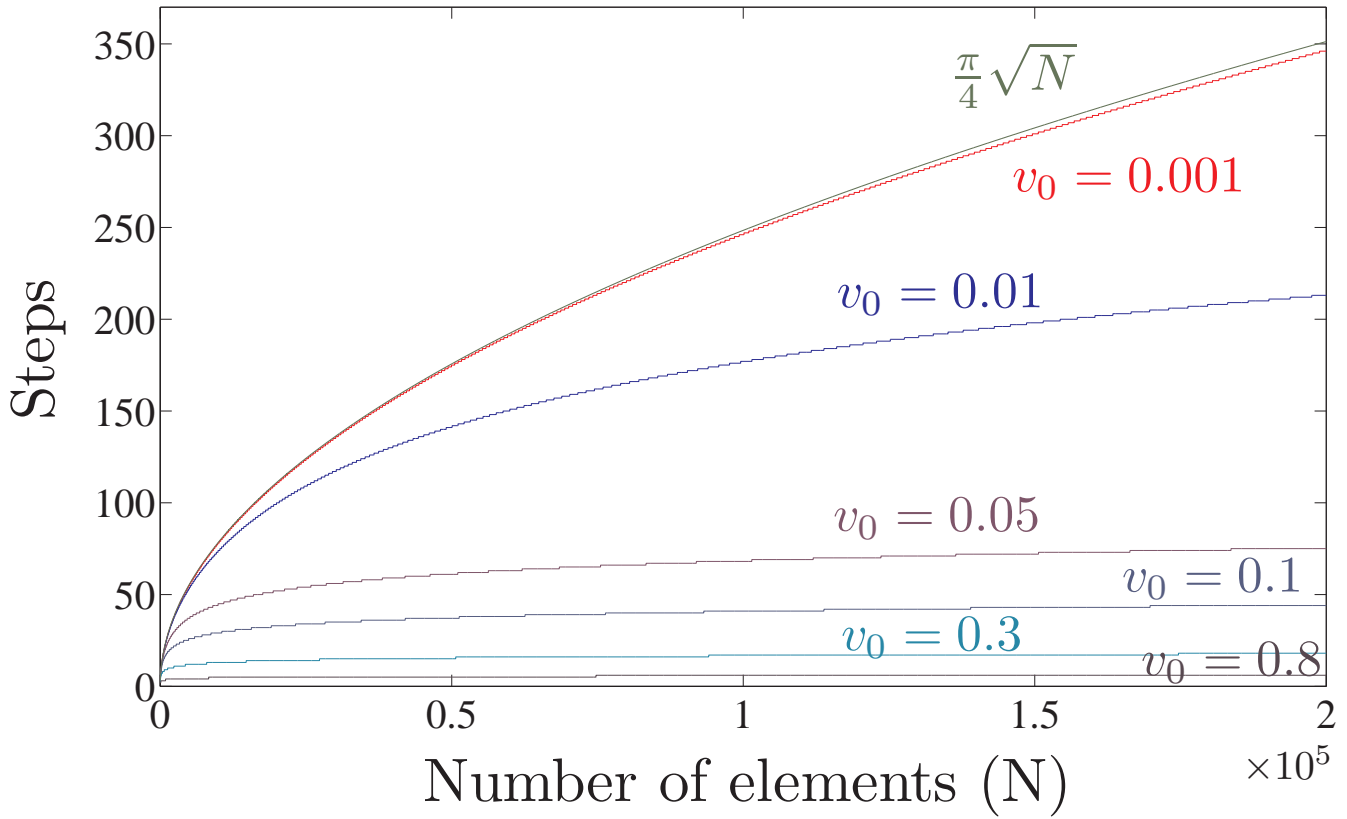}
\caption{\label{steps} Number of iterations before the maximum transfer of energy. As the initial speed increases, the number of collisions needed to transfer most of the kinetic energy to the smaller ball is reduced.}
\end{figure}

Figure \ref{steps} shows the number of steps needed for the maximum energy transfer. As in Grover's algorithm, if we do not stop the procedure, the kinetic energy of the system will alternatively go from the big to the small ball. We have only considered the first maximum of kinetic energy of the small ball (the fastest energy transfer). In the initial state, $\frac{K_2}{K_T}$ is exactly $\frac{1}{N}$. When we stop the procedure, $K_2$ is of the order of $K_T$. The results are given for initial velocities $v_0=0.001, 0.01, 0.05, 0.1, 0.3,$ and $0.8$. The number of steps is plotted along the asymptotic limit of steps for Grover search, $\frac{\pi}{4}\sqrt{N}$ \cite{FG98}. 

The graph shows some representative cases of relativistic Grover energy transfer. The most significant conclusion is that, for this relativistic setting, the procedure is more efficient than in the classical case. The reduction in the number of steps is greater for higher velocities. As expected, for small velocities, the deviation from the classical case is small. However, as we come closer to the speed of light, the number of steps is dramatically reduced. For a high enough initial velocity $v_0$, the maximum transfer can occur even in a single collision.

We can estimate the initial velocity needed for this single step transfer from four-momentum conservation. In our only collision, equations (\ref{consrele}) and (\ref{consrelmom}) become
\begin{eqnarray}
N\gamma_0&=&M+\gamma_2'\\
N\gamma_0v_0-2\gamma_0v_0&=&\gamma_2'v_2'.
\end{eqnarray}
The left hand side shows initial energy and momentum and the right hand side corresponds to a perfect transfer in which the big ball stops ($v_1'=0$) and all the kinetic energy goes to the small ball. If we take $v_0=1-x$ and $v_2'=1-y$, for $x,y<<1$ and assume $N>>1$, we can subtract the equations and find a single step velocity $v_0^{ss}=1-\frac{2}{N^2}$. Numerical simulations show maximum energy transfer in a single collision for a given $N$ and values of $v_0$ around $v_0^{ss}$.

Even for very small initial velocities, there will be a value of $N$ for which relativistic effects will become relevant. This can be best noticed looking at the graph of $v_0=0.001$ in Fig. \ref{steps}. For moderate values of $M$ the deviation from classical is negligible. However, for high values of $M$ the curve starts to separate from the classical prediction. 

We can explain the separation looking into the Taylor expansion of $K_1$ for small initial velocities, $K_1=\frac{1}{2}Mv_0^2+\frac{3}{8}Mv_0^4+O(v_0^6)$. In most problems with a small $v_0$, we could neglect all the terms but the classical kinetic energy. However, for high values of $M$, the correction term can be comparable to the other relevant energy of the problem, $K_2$. We have taken $K_1$ for the initial state, when $v_1=v_0$, as a limit. During Grover energy transfer, $v_1$ becomes smaller at every step as the big mass losses energy. The initial state gives us the maximum deviation of $K_1$ from the classical approximation. 

For a given problem (a particular value of $N$), we can define a breakpoint initial velocity $v_0^b$ at the value at which the first correction term of $K_1$ is equal to the classical approximation of $K_2$, $\frac{3}{8}Mv_{0b}^4=\frac{1}{2}v_{0b}^2$. This happens when $v_0^b=\frac{2}{\sqrt{3M}}$. Velocities much below $v_0^b$ will follow the classical Grover energy transfer, while velocities well above $v_{0b}$ will show a clear deviation. Similarly, for a given $v_0$, we can define a breakpoint $M^b=\frac{4}{3v_0^2}$. This estimation corresponds well to the behaviour shown in Fig. \ref{steps} and is best noticed for $v_0=0.001$ (with $N_b\approx 1.3\cdot 10^6$) and $v_0=0.01$ (with $N_b\approx1.3\cdot 10^4$). 

\subsection{Transferred fraction of energy}
\begin{figure}
\includegraphics[width=\linewidth]{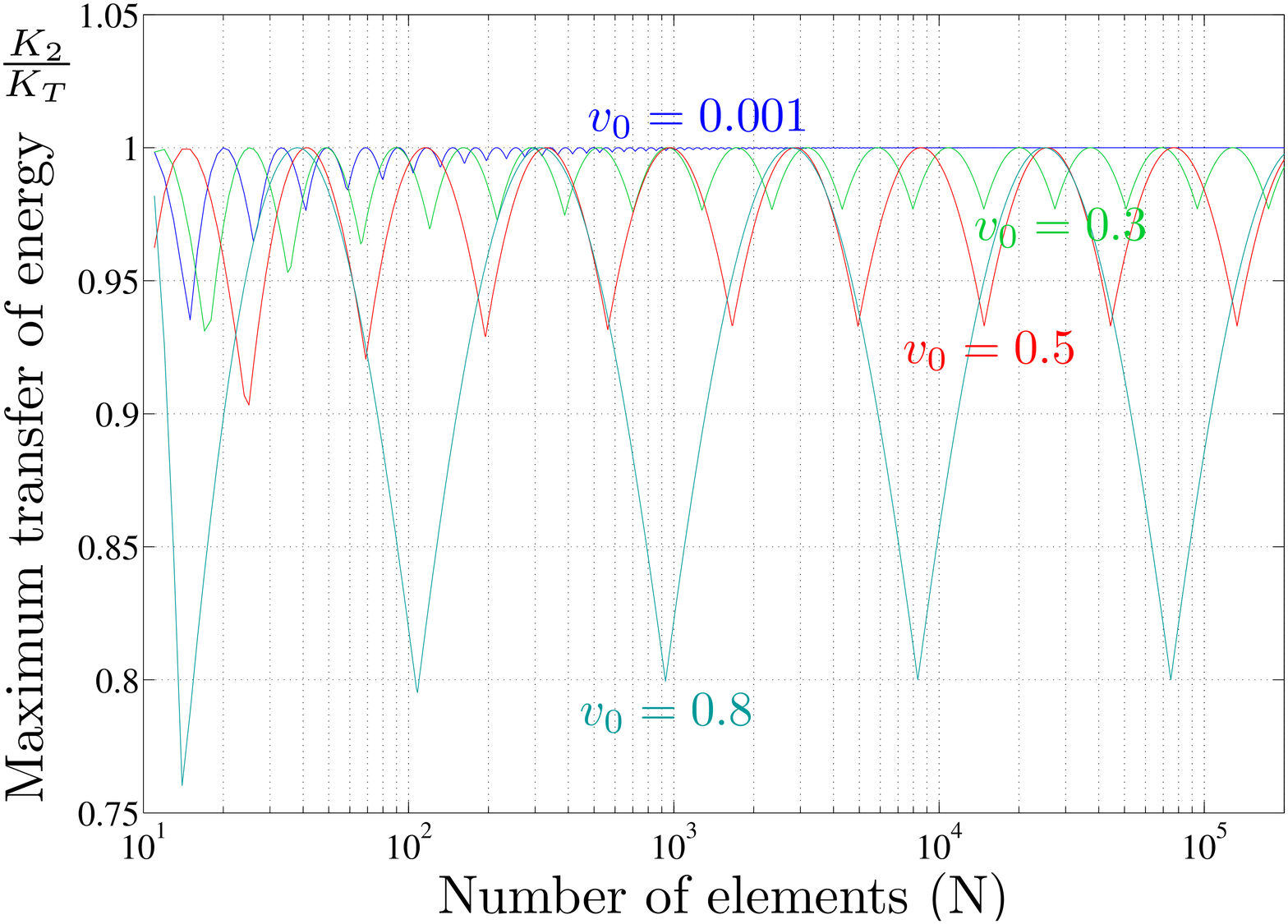}
\caption{\label{Amps} Fraction of the total kinetic energy $K_T$ that the small ball has at the end of the Grover energy transfer.}
\end{figure}

We can also look at the fraction of the total energy that is transferred at the end of the procedure. Figure \ref{Amps} shows that, while in the final step the energy of the small ball is still in the order of the total energy, there can be a significant variation for different initial conditions. 

This behaviour is not exclusive of the relativistic case. In the quantum and classical Grover algorithms there is also a point at which an additional collision cannot increase the energy of the marked element any more. In the Grover procedure, energy is transferred in discrete steps. The final energy of the small ball comes from adding these steps and cannot exceed the total energy. Steps with an energy transfer above the difference $K_T-K_2$ are only allowed if they take energy from the small to the big ball.

This can be compared to the end of a Game of the Goose game. If the die roll exceeds the exact value needed to reach home, the piece rebounds and moves back the excess number of positions. 

In a Grover procedure, a total or almost total transfer can be achieved if the discrete transfer steps carry a low energy and we have a greater precision. When each step transfers a higher energy/probability, as it happens in both classical and quantum Grover algorithms for small values of $N$, the initial value of $N$ will determine the maximum final amplitued. The difference between the maximum transfer values vanishes as $N$ grows and the steps are more precise. For the same reason, the varying maximum amplitude effect becomes more marked at relativistic speeds. When the initial velocity is high, the energy transferred at each step is greater and we have a rougher approximation to an exact transfer of all of the kinetic energy. 

This means that choosing an appropriate pair of values $v_0$ and $N$ can improve the efficiency. There will be some initial values for which we can obtain a perfect fit to the maximum energy (an exact die roll). 

\section{ANALYSIS AND FUTURE WORK}
We have presented results showing that the effective procedure for Grover energy transfer can be more efficient when taken from a classical to a relativistic setting. Table \ref{Comp} gives a comparison between the time and space resources required by the different Grover procedures in the quantum, classical and relativistic cases. The space resources (number of elements) involved in the classical and relativistic cases are both of the order of $N$. In the quantum Grover procedure, where we transfer probability amplitude, quantum superposition allows for a greater compactness as $log_2N$ two-level states are enough to encode $N$ values.   

\begin{table}
\caption{\label{Comp}Comparison of the efficiency of equivalent quantum, classical and relativistic Grover procedures.}
\begin{ruledtabular}
\begin{tabular}{ccc}
& Number of elements & Number of steps \\
\hline
\phantom{\Large{(}}Quantum\phantom{\Large{(}} & $log_2(N)$ & $O\left(\sqrt{N}\right)$ \\
Classical & $N$ & $O\left(\sqrt{N}\right)$ \\
Relativistic &$N$ & $1 (v_0\rightarrow 1)$\\
\end{tabular}
\end{ruledtabular}
\end{table}

The time complexity (number of steps) grows as $\sqrt{N}$ in both the quantum and the classical procedures. This bound has been shown to be optimal \cite{Zal99}. However, the nonlinearities that appear in the relativistic version of the procedure alter the behaviour and allow a faster transfer. For velocities close to the speed of light the transfer can even be done in a single step.

These results show that there are physical algorithmic procedures that are more efficient in situations where relativistic effects apply. This suggests a new field of Relativistic Computation, parallel to Quantum Computation. In this particular case, direct application is doubtful, although the Grover energy transfer procedure could be of interest in particle colliders, as long as elastic collisions could be achieved. 

Nevertheless, the procedure could be implemented with other nonlinear analogues. There have already benn analog simulators of interesting relativistic effects like Hawking radiation in optical fibre systems \cite{PKR08}. Similarly, relativistic Grover energy transfer could be simulated with optical nonlinear systems inside optical fibre or microring resonators. This also opens the door to a new line of research on nonlinear Grover procedures in different systems like, for instance, a series of electronical oscillators with nonlinear coupling.

There are also many possible extensions to the presented Grover transfer procedure. The most obvious addition is checking the procedure's efficiency when we have more than one marked object. In a quantum Grover algorithm with $k$ targets, the number of steps can be divided by a factor of $\sqrt{k}$ \cite{BBH99}. We have already performed preliminary simulations for more than one marked object and, while there is an improvement for high velocities when compared to the classical transfer, the advantage reduces as the number of marked objects increases. A detailed analysis of this data will be presented elsewhere.

Additionally, it would be desirable to have at least a rough approximation to the number of steps the procedure requires for each initial velocity. One of the greatest challenges in dealing with relativistic Grover energy transfer is finding intuitive explanations for the results. The nonlinearity of the equations obscures an easy interpretation. Our work in progress includes a search for closed expressions for the assymptotic limit of steps and combinations of $v_0$ and $N$ that optimise the transfer. Curiously, the classical analogue of Grover algorithm gives a simple physical argument of why the optimal number of steps should be of the order of $\sqrt{N}$ \cite{GS02}. 

Relatedly, it is still to be proven if the relativistic procedure is also optimal. The fact that we recover the $\sqrt{N}$ bound the at the classical limit ($v_0\rightarrow 0$) suggests that it is, but there could be a different behaviour for higher velocities. 

Finally, in a more general programme for Relativistic Information, it appears the exciting possibility of adapting, or even creating, new different algorithms considering both special and general relativity.

\section{Acknowledgements}
This work has been funded by JCyL Grant No. VA001A08 and by Programa Jos\'e Castillejo of the Spanish MICINN Grant Ref. JC2009-00271.


\newcommand{\noopsort}[1]{} \newcommand{\printfirst}[2]{#1}
  \newcommand{\singleletter}[1]{#1} \newcommand{\switchargs}[2]{#2#1}

\end{document}